# CoBWeb: a user-friendly web application to estimate causal treatment effects from observational data using multiple algorithms

Andreas Markoulidakis[*,1,2], *Peter Holmans*[3], *Philip Pallmann*[2], AMonica Busse[2], Beth − Ann Griffin[4]

[1] *School of Medicine, Cardiff University, Cardiff, UK*

[2] *Centre for Trials Research, Cardiff University, Cardiff, UK*

[3] *Division of Psychological Medicine and Clinical Neurosciences, School of Medicine, Cardiff University, Cardiff, UK*

[4] *RAND Corporation, Arlington, VA, USA*

[*] *Corresponding author email: MarkoulidakisA@cardiff.ac.uk*

**Background/aims:** While randomized controlled trials are the gold standard for measuring causal effects, robust conclusions about causal relationships can be obtained using data from observational studies if proper statistical techniques are used to account for the imbalance of pretreatment confounders across groups. Propensity score (PS) and balance weighting are useful techniques that aim to reduce the observed imbalances between treatment groups by weighting the groups to be as similar as possible with respect to observed confounders.

**Methods:** We have created CoBWeb, a free and easy-to-use web application for the estimation of causal treatment effects from observational data, using PS and balancing weights to control for confounding bias. CoBWeb uses multiple algorithms to estimate the PS and balancing weights, to allow for more flexible relations between the treatment indicator and the observed confounders (as different algorithms make different (or no) assumption about the structural relationship between the treatment covariate and the confounders). The optimal algorithm can be chosen by selecting the one that achieves the best trade-off between balance and effective sample size.

**Results:** CoBWeb follows all the key steps required for robust estimation of the causal treatment effect from observational study data and includes sensitivity analysis of the potential impact of unobserved confounders. We illustrate the practical use of the app using a dataset derived from a study of an intervention for adolescents with substance use disorder, which is available for users within the app environment.

**Conclusion:** CoBWeb is intended to enable non-specialists to understand and apply all the key steps required to perform robust estimation of causal treatment effects using observational data.

## Introduction

Propensity Score (PS) and balancing weights are deployed in the estimation of the causal treatment effect in observational studies, to control for confounding bias due to the non-random allocation of the individuals in the treatment groups, as opposed to a randomized controlled trial (RCT). In the latter, individuals are assigned randomly to treatment or control group, thus any difference in the outcomes of the groups at the end of the study is assumed to rely exclusively on the intervention/treatment.

The main goal of the majority of observational studies is the estimation of the causal treatment effect (Holland, 1986). The causal treatment effect of a treatment for each individual is defined as the difference in the outcome for an individual had they received that treatment compared to the outcome had they not received it (Markoulidakis, Taiyari, et al., 2021). Since we are able to measure the state of an individual in only one case (Rosenbaum & Rubin, 1983) (either received the treatment or not) we use data from observational studies, where two (or more) groups receive different treatments over time, and then we estimate the causal treatment effect from the difference in the outcomes of the groups. If we do not adjust for confounding, the estimate will almost certainly be biased.

The PS (Rubin, 1977) is the probability of an individual's allocation to a specific treatment group, given their observed baseline (pre-treatment) characteristics. PS are used to create comparable treatment groups, either by weighting (Hernán et al., 2000; Robins et al., 2000), matching, adjusting or stratifying on the PS. PS methods attempt to reduce the bias in the estimation of the causal treatment effect due to the observed confounders (Markoulidakis, Taiyari, et al., 2021), by minimization of the imbalance of known and observed confounders between the treatment groups.

Robust causal estimation requires multiple steps which are only implemented in specialist software — often requiring the use of multiple packages — and therefore not readily accessible to non-specialists. In this article, we introduce CoBWeb (Covariate Balancing & Weighting Web App), a free and easy-to-use web tool for estimation of the causal treatment effect from observational data, using multiple PS and balancing weights algorithms, thus maximizing the chance of finding a set of weights resulting in good covariate balance across groups. The app follows the key steps described in (Markoulidakis, Taiyari, et al., 2021) for the robust and unbiased estimation of the causal treatment effect, and provides a user-friendly way for non-specialists to apply these state-of-the-art statistical techniques. The app is freely available at

Other recently published apps for estimating causal treatment effects include:

- The *TWANG shiny* app (Ridgeway et al., 2017), which uses only one algorithm for the estimation of causal treatment effect based on one PS weighting algorithm , and omits the steps of overlap/outliers assessment and sensitivity analysis.
- *TippingSens* (Haensch et al., 2020) and *Propensity Score Matching* (Demir et al., 2020). These apps perform PS matching, rather than balancing and weighting. Furthermore, they do not assess the overlap in covariate values between groups, or the presence of outliers, both of which are necessary for accurate estimation of the causal treatment effect.

Other tools for estimation of causal treatment effects are available in the form of packages/extensions of programming languages (like R (Bryer et al., 2017, 2018; Seibold et al., 2019), SAS, STATA (McCarthy et al., 2014), SPSS (Levesque & others, 2007)),which often require specialist programming skills and/or the purchase of expensive licenses. To the best of our knowledge, none of them provides all the key steps (Markoulidakis, Taiyari, et al., 2021). This may require users to choose optimal analysis parameters in several packages, a challenging task for non-experts.

The remainder of this article is organised as follows. Section 2 describes the basic principles/steps that one should follow to obtain robust and unbiased inference from observational data (2.1), and the software used to implement these (2.2). Section 3, describes the example data we use to illustrate the features of the app, before section 4 presents a step by step illustration of the app modules. Section 5 concludes.

# Methods

## Statistical Methods

CoBWeb follows closely the steps described in detail in the related tutorial paper on estimating robust causal treatment effects from observational data (Markoulidakis, Taiyari, et al., 2021). These steps comprise (1) choice of the estimand of interest, (2) evaluation of the sample for any obvious outliers and overlap concerns among the treatment groups (using both summary statistics and graphical tools), (3) computation of PS and balancing weights using multiple algorithms, (4) assessment of balance achieved on the baseline confounders using the weights, (5) modeling of the outcome and estimation of the treatment effect, and (6) sensitivity analysis using innovative tools to assess the robustness of the estimation regarding both the value of treatment effect and its statistical significance to unobserved confounders. We decribe each in brief detail here.

### **Step 1.** Choose which estimand one is interested in (ATE, ATT, ATC).

The most commonly used causal treatment effects are: 1. the *Average Treatment Effect on the Entire population* (ATE), 2. the *Average Treatment Effect on the Treated population* (ATT), and 3. the *Average Treatment Effect on Control population* (ATC). ATE allows one to understand the average causal treatment effect for the entire population of individuals in both the treatment and control groups, while ATT (ATC) allows one to understand the effect of the treatment among only individuals like those in the treatment (control) group, respectively.

### **Step 2.** Assess sample for lack of overlap in confounders

and adjust as needed. Lack of overlap in confounder values between the treatment and control group violates the assumptions of Rubin's Causal Model (RCM), that each individual should have a positive probability of assignment to each treatment group (Rosenbaum &

Rubin, 1983). Lack of overlap and/or the presence of outliers could affect the weighting procedure by allowing extreme weights to occur (Markoulidakis, Taiyari, et al., 2021). If one retains outliers in the sample, then the algorithms computing PS and balancing weights could assign extreme weights to these individuals, resulting in an effect estimate which could be driven by a small number of observations (and potential bias). CoBWeb allows the user to examine the summary statistics of every confounder, as well as their density plots, providing the option to trim outliers for either upper or lower tail, or both, for one or more confounders at a time.

**Step 3.** Estimate the propensity score or balancing weights using multiple methods

Even though there have been several articles comparing PS and balancing weighting methods (for example, (Abdia et al., 2017; Griffin et al., 2017; Mao et al., 2019; Setodji et al., 2017; Setoguchi et al., 2008)), none of them has identified a method that achieves superior balance to all others in all practically relevant situations, and different methods achieve better balance across the baseline characteristics under different settings — depending on the structure of the given data set (sample size, number of covariates to be balanced, true underlying form of the treatment assignment model, etc.). Therefore, it is advantageous to use multiple methods to estimate the weights in order to optimise balance. CoBWeb computes PS and balancing weights using logistic regression (Agresti, 2018) (LR), the Covariate Balance Propensity Score method (Imai & Ratkovic, 2014) (CBPS), the Generalized Boosted Model) (Huang MY Vegetabile B, n.d.; Ridgeway, 1999) (GBM) and Entropy Balancing (Hainmueller, 2012) (EB).

**Step 4.** Assess balance and effective sample size for all methods and choose the best one for outcome analysis.

Balance (or comparability) among the groups needs to be assessed in the weighted treatment groups. The theory suggests that balance should be obtained on the full multivariate distribution of the observed confounders after one applies the weights (Markoulidakis, Taiyari, et al., 2021), which is rarely assessed in practice. CoBWeb reports both the *standardized mean difference* (SMD) and the *Kolmogorov-Smirnov statistic* (KS) as a way to assess how comparable the two treatment groups are in the baseline characteristics. The former measures the difference between the mean values of the baseline confounders for the two (weighted) treatment groups, while the latter allows us to assess balance also in the tails of the distributions for a given confounder (Markoulidakis, Holmans, et al., 2021). These metrics are commonly used in the literature(Austin et al., 2007; Franklin et al., 2014; Gail & Green, 1976; Griffin et al., 2017; Mlcoch et al., 2019; Setodji et al., 2017; Setoguchi et al., 2008; Zhang et al., 2019). We recommend that adequate balance is achieved when all covariates reports a KS statistic bellow 0.1 across the treatment groups (Markoulidakis, Holmans, et al., 2021).

Additionally, the app computes the *effective sample size* (ESS) of the weighted data for each algorithm. ESS is a useful metric to understand the impact of the weighting on the original sample, as it is a measure of the precision in estimating the treatment effect — ESS expresses the magnitude of the remaining sample size after the weighting procedure (the weights shrink the sample size), and serves as a selection criterion when multiple algorithms achieve adequate balance — in such cases the one with the higher ESS should be selected (Markoulidakis, Taiyari, et al., 2021).

### **Step 5.** Model the outcome and estimate the causal treatment effect.

Instead of estimating the causal treatment effect as the difference between the weighted means of the outcome values, CoBWeb combines the weights with a multivariable regression adjustment that ideally includes all of the observed confounders used in the estimation of the weights (Austin, 2011; Ridgeway et al., 2017), — this approach yields the so-called *doubly robust estimator* of the causal treatment effect (Bang & Robins, 2005; Chattopadhyay et al., 2020; Kang et al., 2007; Zhao & Percival, 2016). Using this estimator, the estimated treatment effect is consistent (asymptotically unbiased) so long as one part of the doubly robust model is correct (either the PS weight model or the multivariable outcome model).

### **Step 6.** Assess sensitivity of the results to unobserved confounding.

Weighted analyses are based on the assumption that all potential confounders have been measured and included in the computation of PS and balancing weights. In practice, it is impossible to know if there are any unobserved confounders. Thus, it is crucial to perform sensitivity analysis, in order to evaluate the robustness of the causal treatment effect estimation (both the value of the estimation and its statistical significance) to unobserved confounders (Griffin et al., 2020).

## Software Implementation

CoBWeb was written in the *R* (Team, 2018) programming language, using the extension package *shiny* (Chang et al., 2017), which provides a framework for building interactive web applications.

To compute the PS and balancing weights, the app uses the following packages: 1. *CBPS* (Ratkovic et al., 2013) for the computation of the PS using CBPS algorithm, controlling for the first $m = 1,2,3$ moments; 2. *twang* (Ridgeway et al., 2020) to compute the PS via the GBM algorithm, using the mean SMD ($GBM_{ES}$) and the maximum KS ($GBM_{KS}$) as stopping rules; 3. *entbal* (Vegetabile et al., 2021), to compute the balancing weights, using entropy balancing, controlling for the first $m = 1,2,3$ moments.

*cobalt* (Fong et al., 2019; Greifer, 2020) and *survey* (Lumley, 2020) packages are used to compute the balance measures (SMD and KS statistics) used to assess the balance between the treatment groups after the weighting.

*rmarkdown* (Xie et al., 2018) is used to produce the .pdf documents, while *xtable* (Dahl et al., 2019) and *tinytex* (Xie, 2019) are used to create the tables included on the .pdf documents.

*OVtool* (Pane et al., n.d.) is used to produce the sensitivity analysis plot (the last step of the analysis), while the remaining plots are produced using *ggplot2* (Wickham et al., 2016) and *GGally* (Schloerke et al., 2018).

## Data

Throughout the demonstration of the app, we will utilize a synthetic data-set based on a longitudinal observational study data on adolescents receiving substance use disorder (SUD) treatment, who were administered the Global Appraisal of Individual Needs (GAIN) biopsychosocial assessment instrument (Dennis et al., 2003) regularly. The synthetic data-set is available for the users of the *CoBWeb* app on the *Data* module. Synthetic data mimic the statistical properties of the original data, but the observations are simulated, and thus not real — we don not provide the real data, due to data protection issues and to protect study participants identities.

In our edition of the data, there are 2000 individuals in each group, receiving two different treatments - *treatment* and *control*, respectively. More information about the data can be found in (Dennis et al., 2003), and a brief definition of the covariates available on the synthetic data-set is available for the users in the *Data* module — button *"About the Example Data-set"*.

## Results

### *The CoBWeb App*

The CoBWeb app has 7 main sections: 1. Welcome!, 2. Data, 3. Model Set-Up, 4. Overlap Assessment, 5. Balance Evaluation, 6. Outcome Analysis, and 7. Before you go!. The parts of the app follow the steps described in *section* 2 (Markoulidakis, Taiyari, et al., 2021). We will keep maintaining the functionality of the web app (including bug fixes and possibly adding in new functionality); hence, the appearance and functionality of CoBWeb may change slightly as it evolves.

### *Welcome!*

The "Welcome" page is the initial module of the app, where general information about the goal of the app is provided, alongside a link to the article (Markoulidakis, Taiyari, et al., 2021) that describes the steps that one should take to obtain a robust estimation of the causal treatment effect using observational data.

### *Data*

The "Data" module, provides the user with the option to upload their data, or to utilize the example data-set, to understand the features of the app, and to become familiar with the causal treatment effect analysis. Throughout this tutorial, we will utilize the example data-set described above (just tap on the option *Load Example Data* — left yellow box, Figure 1). Once the data are loaded (either example data, or uploading data via computer browser), a table with summary statistics (including mean and standard deviation per covariate) will become available, as well as the first few rows or the raw data (the number of rows is controlled by the user via the option *Number of observations to view*). More information about our example data-set can be found by tapping the button *About this Example Data-set*, which is also described in the tutorial (Markoulidakis, Taiyari, et al., 2021).

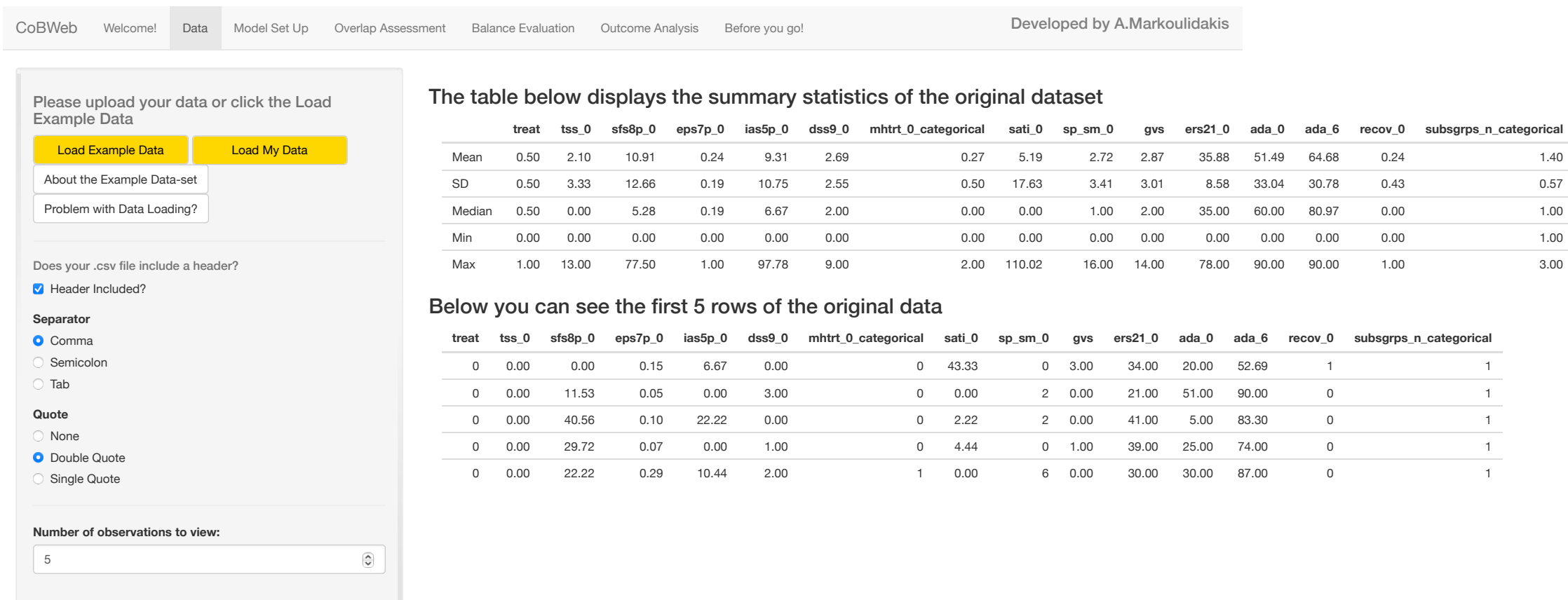

| | treat | tss_0 | sfs8p_0 | eps7p_0 | ias5p_0 | dss9_0 | mhtrt_0_categorical | sati_0 | sp_sm_0 | gvs | ers21_0 | ada_0 | ada_6 | recov_0 | subsgrps_n_categorical |
|---|---|---|---|---|---|---|---|---|---|---|---|---|---|---|---|
| Mean | 0.50 | 2.10 | 10.91 | 0.24 | 9.31 | 2.69 | 0.27 | 5.19 | 2.72 | 2.87 | 35.88 | 51.49 | 64.68 | 0.24 | 1.40 |
| SD | 0.50 | 3.33 | 12.66 | 0.19 | 10.75 | 2.55 | 0.50 | 17.63 | 3.41 | 3.01 | 8.58 | 33.04 | 30.78 | 0.43 | 0.57 |
| Median | 0.50 | 0.00 | 5.28 | 0.19 | 6.67 | 2.00 | 0.00 | 0.00 | 1.00 | 2.00 | 35.00 | 60.00 | 80.97 | 0.00 | 1.00 |
| Min | 0.00 | 0.00 | 0.00 | 0.00 | 0.00 | 0.00 | 0.00 | 0.00 | 0.00 | 0.00 | 0.00 | 0.00 | 0.00 | 0.00 | 1.00 |
| Max | 1.00 | 13.00 | 77.50 | 1.00 | 97.78 | 9.00 | 2.00 | 110.02 | 16.00 | 14.00 | 78.00 | 90.00 | 90.00 | 1.00 | 3.00 |

| treat | tss_0 | sfs8p_0 | eps7p_0 | ias5p_0 | dss9_0 | mhtrt_0_categorical | sati_0 | sp_sm_0 | gvs | ers21_0 | ada_0 | ada_6 | recov_0 | subsgrps_n_categorical |
|---|---|---|---|---|---|---|---|---|---|---|---|---|---|---|
| 0 | 0.00 | 0.00 | 0.15 | 6.67 | 0.00 | 0 | 43.33 | 0 | 3.00 | 34.00 | 20.00 | 52.69 | 1 | 1 |
| 0 | 0.00 | 11.53 | 0.05 | 0.00 | 3.00 | 0 | 0.00 | 2 | 0.00 | 21.00 | 51.00 | 90.00 | 0 | 1 |
| 0 | 0.00 | 40.56 | 0.10 | 22.22 | 0.00 | 0 | 2.22 | 2 | 0.00 | 41.00 | 5.00 | 83.30 | 0 | 1 |
| 0 | 0.00 | 29.72 | 0.07 | 0.00 | 1.00 | 0 | 4.44 | 0 | 1.00 | 39.00 | 25.00 | 74.00 | 0 | 1 |
| 0 | 0.00 | 22.22 | 0.29 | 10.44 | 2.00 | 1 | 0.00 | 6 | 0.00 | 30.00 | 30.00 | 87.00 | 0 | 1 |

### *Model Set-Up*

The "Model Set-Up" module (Figure 2) is where the user will name the variables that are the *Treatment Indicator* (*treat* variable in our example data, and how the control (0) and treatment (1) group are represented on the treatment indicator variable. These variables can be either numerical $(0,1,2,...)$ or categorical $(A, B, ..)$. The app will show an error message on the left panel if there are more than two levels on the treatment indicator variable selected — the current version of the app considers only two treatment groups. Then, the user should declare the *Outcome* variable (*ada*_6 for the example data-set), choosing from a list showing all the available data-set variables. Next, *Binary & Continuous Confounders* and *Categorical Confounders* should be chosen. The categorical confounders with more than two categories, will be transformed to dummies, such that a dummy variable will be created for each original category, with the user choosing which level of each categorical confounder will be the baseline. Finally, the user must choose the treatment effect of interest. There are two options, $ATE$ and $ATT$, but, one could perform the analysis for the estimation of the $ATC$, by choosing $ATT$ as the treatment effect of interest, and flipping the control and treatment group values. Every selection made will be shown on the right side of the panel, for verification of the models.

In the example data-set, $mhtrt_0_categorical$ and $subsgrps_n_categorical$ are categorical covariates. All the other covariates (except *treat*) are continuous numerical.

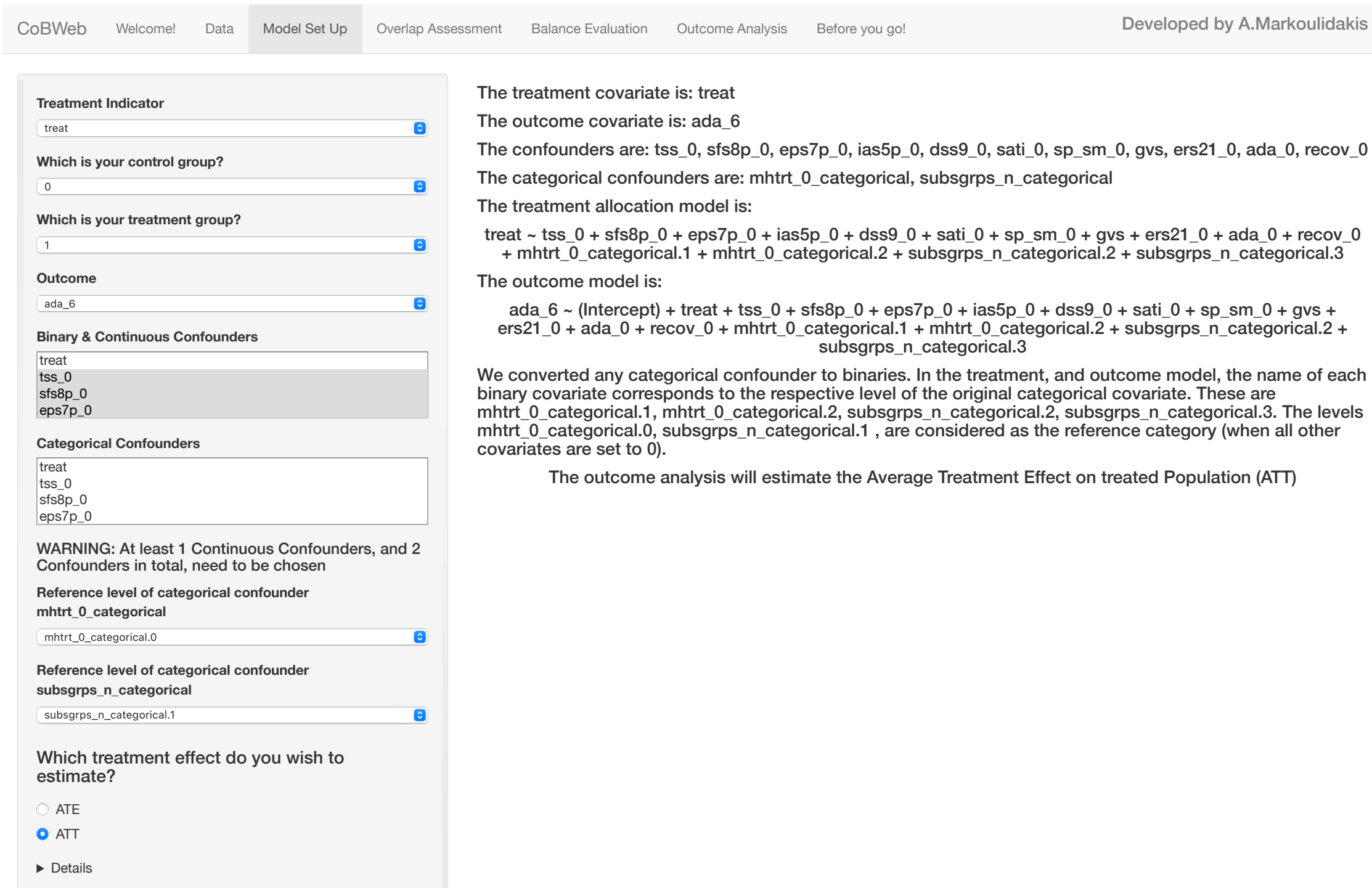

## *Overlap Assessment*

The "Overlap Assessment" module consists of three sub-modules. The first shows tables of summary statistics for the control and treatment groups (*Treatment Groups Summaries* ). The second shows density plots of each confounder for the two groups, allowing lack of overlap to be detected (*Dealing with Outliers* — Figure 3). The third sub-module shows revised summary statistics of the treatment groups, after removal of outliers and observations with missing values (*Final Data Treatment Groups Summaries* ). The second sub-module (*Dealing with Outliers*) gives the user the option to remove outliers from the upper, lower or both tails of the distribution of each confounder, to correct lack of overlap for the treatment groups. When the user selects outliers based on the observations of one confounder, the observations are removed from the entire data-set, and thus all density plots are updated. This can help resolve overlap issues on other confounders if one observation reports problematic values on several measures.

In our example dataset, the treatment groups overlap across all confounders, thus there is no need to remove any outliers.

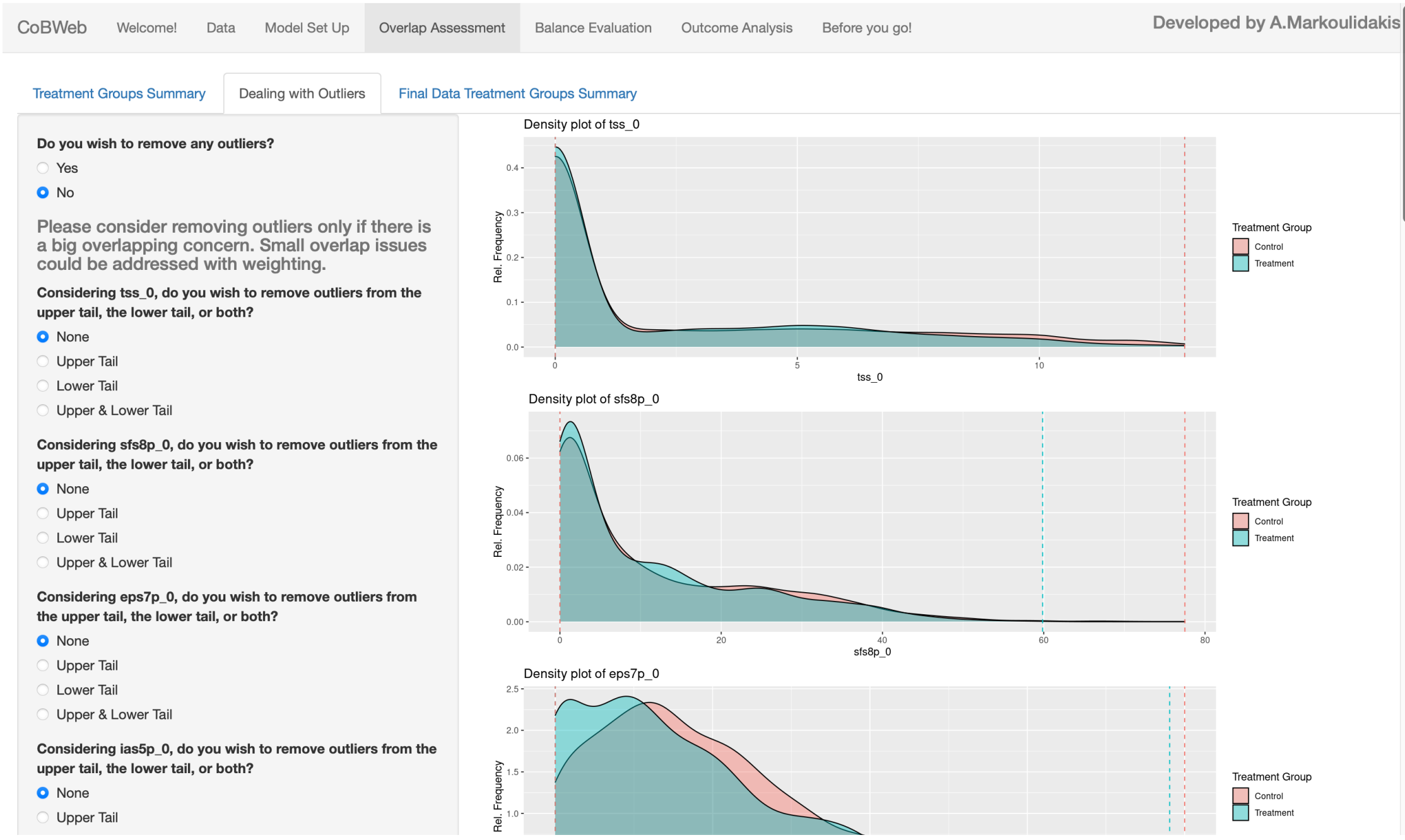

## *Balance Evaluation*

The "Balance Evaluation" module (*Figure* 4) reports the SMD (first table) and the KS statistic (second table) values for each confounder, as well as the mean and maximum values (the last two lines of each of the first two tables), for each of the algorithms used to obtain PS and balancing weights. Finally, the third (lower) table, reports the ESS in absolute value, and as a percentage of the original size, for each algorithm.

PS and balancing weights are used to weight the treatment groups, such that the two groups are balanced, with respect to all known and measured covariates, after weighting. The treatment groups are considered to be well balanced if the maximum SMD and the maximum KS values are both less than 0.1 — this means the SMD and KS values are below 0.1 for every confounder. If the maximum KS value for an algorithm is above the threshold of 0.1, this means that the treatment groups are not balanced (Markoulidakis, Holmans, et al., 2021). We suggest that, for the final outcome analysis, users pick the PS and balancing weights algorithm that achieves the best balance in terms of maximum KS (that is, the lowest value of maximum KS reported) (Markoulidakis, Taiyari, et al., 2021). If multiple algorithms achieve similar performance in terms of balance, the algorithm with the higher ESS value should be chosen (Markoulidakis, Taiyari, et al., 2021).

In the example data-set, $\text{GBM}_{ES}$, and $\text{GBM}_{KS}$ achieve the best balance (maximum KS value equal to 0.02 — last row of second table, *Figure* 4). Of these, $\text{GBM}_{ES}$ achieves the highest ESS value, thus this algorithm is chosen for the outcome analysis. It should be noted that logistic regression gives the highest value of ESS (the weighted data retains 94% of the size of the original data), and since the weights produced from this algorithm balance the treatment groups (maximum KS value 0.06), this algorithm could also have been chosen for the outcome analysis. The EB algorithms report an ESS of 2002, much lower than the other methods. This is due to the extreme weights (Markoulidakis, Holmans, et al., 2021) produced by the EB algorithm (set to 0) — since we estimate the ATT, we compute weights only for the control group, thus the ESS of the treatment group remains untouched (equal to 2000), while that of the control group is reduced from 2000 to 2 — EB often computes extreme balancing weights in the attempt to achieve a perfect balance on the predefined moments.

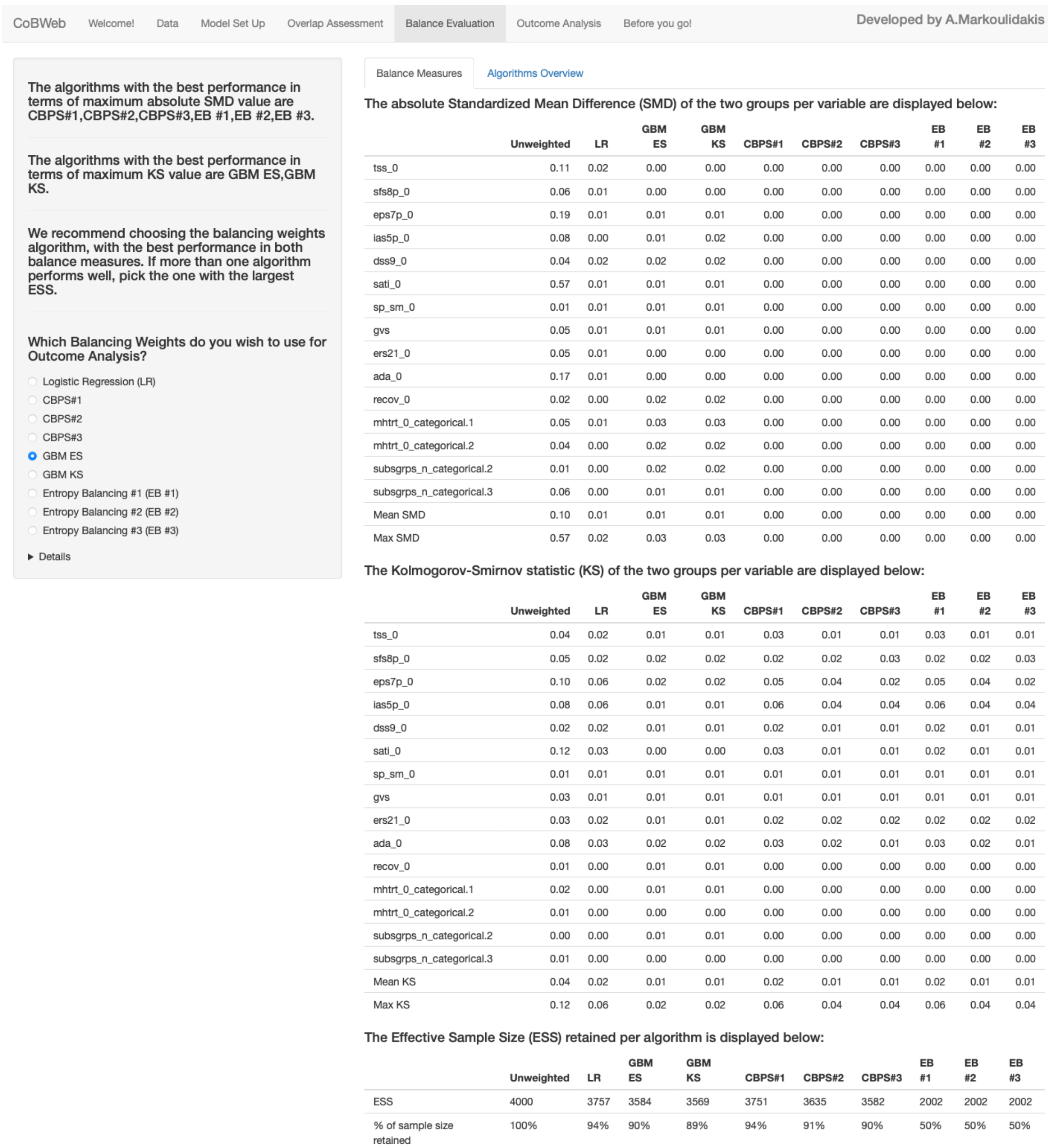

The absolute Standardized Mean Difference (SMD) of the two groups per variable are displayed below:

| | Unweighted | LR | GBM ES | GBM KS | CBPS#1 | CBPS#2 | CBPS#3 | EB #1 | EB #2 | EB #3 |
|---|---|---|---|---|---|---|---|---|---|---|
| tss_0 | 0.11 | 0.02 | 0.00 | 0.00 | 0.00 | 0.00 | 0.00 | 0.00 | 0.00 | 0.00 |
| sfs8p_0 | 0.06 | 0.01 | 0.00 | 0.00 | 0.00 | 0.00 | 0.00 | 0.00 | 0.00 | 0.00 |
| eps7p_0 | 0.19 | 0.01 | 0.01 | 0.01 | 0.00 | 0.00 | 0.00 | 0.00 | 0.00 | 0.00 |
| ias5p_0 | 0.08 | 0.00 | 0.01 | 0.02 | 0.00 | 0.00 | 0.00 | 0.00 | 0.00 | 0.00 |
| dss9_0 | 0.04 | 0.02 | 0.02 | 0.02 | 0.00 | 0.00 | 0.00 | 0.00 | 0.00 | 0.00 |
| sati_0 | 0.57 | 0.01 | 0.01 | 0.01 | 0.00 | 0.00 | 0.00 | 0.00 | 0.00 | 0.00 |
| sp_sm_0 | 0.01 | 0.01 | 0.01 | 0.01 | 0.00 | 0.00 | 0.00 | 0.00 | 0.00 | 0.00 |
| gvs | 0.05 | 0.01 | 0.01 | 0.01 | 0.00 | 0.00 | 0.00 | 0.00 | 0.00 | 0.00 |
| ers21_0 | 0.05 | 0.01 | 0.00 | 0.00 | 0.00 | 0.00 | 0.00 | 0.00 | 0.00 | 0.00 |
| ada_0 | 0.17 | 0.01 | 0.00 | 0.00 | 0.00 | 0.00 | 0.00 | 0.00 | 0.00 | 0.00 |
| recov_0 | 0.02 | 0.00 | 0.02 | 0.02 | 0.00 | 0.00 | 0.00 | 0.00 | 0.00 | 0.00 |
| mhtrt_0_categorical.1 | 0.05 | 0.01 | 0.03 | 0.03 | 0.00 | 0.00 | 0.00 | 0.00 | 0.00 | 0.00 |
| mhtrt_0_categorical.2 | 0.04 | 0.00 | 0.02 | 0.02 | 0.00 | 0.00 | 0.00 | 0.00 | 0.00 | 0.00 |
| subsgrps_n_categorical.2 | 0.01 | 0.00 | 0.02 | 0.02 | 0.00 | 0.00 | 0.00 | 0.00 | 0.00 | 0.00 |
| subsgrps_n_categorical.3 | 0.06 | 0.00 | 0.01 | 0.01 | 0.00 | 0.00 | 0.00 | 0.00 | 0.00 | 0.00 |
| Mean SMD | 0.10 | 0.01 | 0.01 | 0.01 | 0.00 | 0.00 | 0.00 | 0.00 | 0.00 | 0.00 |
| Max SMD | 0.57 | 0.02 | 0.03 | 0.03 | 0.00 | 0.00 | 0.00 | 0.00 | 0.00 | 0.00 |

The Kolmogorov-Smirnov statistic (KS) of the two groups per variable are displayed below:

| | Unweighted | LR | GBM ES | GBM KS | CBPS#1 | CBPS#2 | CBPS#3 | EB #1 | EB #2 | EB #3 |
|---|---|---|---|---|---|---|---|---|---|---|
| tss_0 | 0.04 | 0.02 | 0.01 | 0.01 | 0.03 | 0.01 | 0.01 | 0.03 | 0.01 | 0.01 |
| sfs8p_0 | 0.05 | 0.02 | 0.02 | 0.02 | 0.02 | 0.02 | 0.03 | 0.02 | 0.02 | 0.03 |
| eps7p_0 | 0.10 | 0.06 | 0.02 | 0.02 | 0.05 | 0.04 | 0.02 | 0.05 | 0.04 | 0.02 |
| ias5p_0 | 0.08 | 0.06 | 0.01 | 0.01 | 0.06 | 0.04 | 0.04 | 0.06 | 0.04 | 0.04 |
| dss9_0 | 0.02 | 0.02 | 0.01 | 0.01 | 0.02 | 0.01 | 0.01 | 0.02 | 0.01 | 0.01 |
| sati_0 | 0.12 | 0.03 | 0.00 | 0.00 | 0.03 | 0.01 | 0.01 | 0.02 | 0.01 | 0.01 |
| sp_sm_0 | 0.01 | 0.01 | 0.01 | 0.01 | 0.01 | 0.01 | 0.01 | 0.01 | 0.01 | 0.01 |
| gvs | 0.03 | 0.01 | 0.01 | 0.01 | 0.01 | 0.01 | 0.01 | 0.01 | 0.01 | 0.01 |
| ers21_0 | 0.03 | 0.02 | 0.01 | 0.01 | 0.02 | 0.02 | 0.02 | 0.02 | 0.02 | 0.02 |
| ada_0 | 0.08 | 0.03 | 0.02 | 0.02 | 0.03 | 0.02 | 0.01 | 0.03 | 0.02 | 0.01 |
| recov_0 | 0.01 | 0.00 | 0.01 | 0.01 | 0.00 | 0.00 | 0.00 | 0.00 | 0.00 | 0.00 |
| mhtrt_0_categorical.1 | 0.02 | 0.00 | 0.01 | 0.01 | 0.00 | 0.00 | 0.00 | 0.00 | 0.00 | 0.00 |
| mhtrt_0_categorical.2 | 0.01 | 0.00 | 0.00 | 0.00 | 0.00 | 0.00 | 0.00 | 0.00 | 0.00 | 0.00 |
| subsgrps_n_categorical.2 | 0.00 | 0.00 | 0.01 | 0.01 | 0.00 | 0.00 | 0.00 | 0.00 | 0.00 | 0.00 |
| subsgrps_n_categorical.3 | 0.01 | 0.00 | 0.00 | 0.00 | 0.00 | 0.00 | 0.00 | 0.00 | 0.00 | 0.00 |
| Mean KS | 0.04 | 0.02 | 0.01 | 0.01 | 0.02 | 0.01 | 0.01 | 0.02 | 0.01 | 0.01 |
| Max KS | 0.12 | 0.06 | 0.02 | 0.02 | 0.06 | 0.04 | 0.04 | 0.06 | 0.04 | 0.04 |

The Effective Sample Size (ESS) retained per algorithm is displayed below:

| | Unweighted | LR | GBM ES | GBM KS | CBPS#1 | CBPS#2 | CBPS#3 | EB #1 | EB #2 | EB #3 |
|---|---|---|---|---|---|---|---|---|---|---|
| ESS | 4000 | 3757 | 3584 | 3569 | 3751 | 3635 | 3582 | 2002 | 2002 | 2002 |
| % of sample size retained | 100% | 94% | 90% | 89% | 94% | 91% | 90% | 50% | 50% | 50% |

## *Outcome Analysis*

The "Outcome Analysis" module consists of two parts: the *Treatment Effect Estimation* (Figure 5) and *Sensitivity Analysis* (Figure 6). The former reports the estimation of the causal treatment effect (this is the estimate of the *treat* covariate in the regression model). The user has the option to generate a report (as a .pdf document), including the main findings of the analysis, and/or download the refined data (after removing potential outliers) for further analysis. The second sub-module (*Sensitivity Analysis* — Figure 6), will create a graph which helps assess the robustness of the causal treatment effect to confounders that are not measured in the current data-set (Markoulidakis, Taiyari, et al., 2021). The x-axis reports the *Association with Treatment Indicator* (this is the SMD across the groups of a hypothetical new and unobserved confounder), while the y-axis reports the *Absolute Association with Outcome Covariate* — this is the absolute correlation of the unobserved confounder with the outcome covariate. The solid black contour lines represent the value of the treatment effect if the confounder were included in the analysis, while the dotted red contour lines represent the corresponding level of significance (p-value levels). The blue dots are the observed confounders, positioned to represent their association with the treatment indicator (x-axis) and the outcome (y-axis).

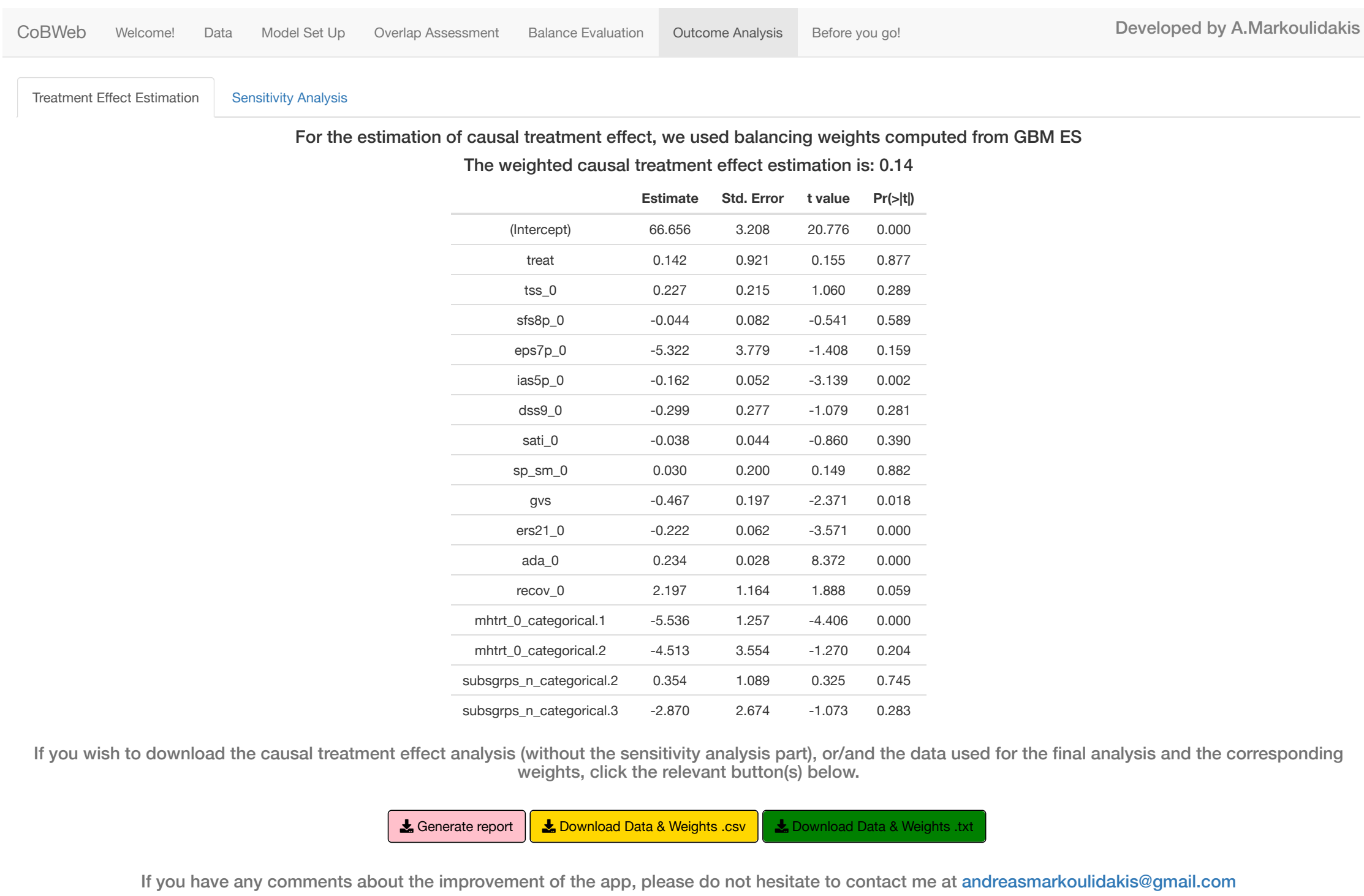

| | Estimate | Std. Error | t value | Pr(>\|t\|) |
|---|---|---|---|---|
| (Intercept) | 66.656 | 3.208 | 20.776 | 0.000 |
| treat | 0.142 | 0.921 | 0.155 | 0.877 |
| tss_0 | 0.227 | 0.215 | 1.060 | 0.289 |
| sfs8p_0 | -0.044 | 0.082 | -0.541 | 0.589 |
| eps7p_0 | -5.322 | 3.779 | -1.408 | 0.159 |
| ias5p_0 | -0.162 | 0.052 | -3.139 | 0.002 |
| dss9_0 | -0.299 | 0.277 | -1.079 | 0.281 |
| sati_0 | -0.038 | 0.044 | -0.860 | 0.390 |
| sp_sm_0 | 0.030 | 0.200 | 0.149 | 0.882 |
| gvs | -0.467 | 0.197 | -2.371 | 0.018 |
| ers21_0 | -0.222 | 0.062 | -3.571 | 0.000 |
| ada_0 | 0.234 | 0.028 | 8.372 | 0.000 |
| recov_0 | 2.197 | 1.164 | 1.888 | 0.059 |
| mhtrt_0_categorical.1 | -5.536 | 1.257 | -4.406 | 0.000 |
| mhtrt_0_categorical.2 | -4.513 | 3.554 | -1.270 | 0.204 |
| subsgrps_n_categorical.2 | 0.354 | 1.089 | 0.325 | 0.745 |
| subsgrps_n_categorical.3 | -2.870 | 2.674 | -1.073 | 0.283 |

In the example dataset, an unobserved confounder with a reported SMD of $-0.6$ between the treatment and control groups, and moderate correlation with the outcome (i.e. correlation coefficient 0.4) would be placed in the extreme top left corner of the plot. The black contour lines corresponding to the treatment effect estimates indicate that the inclusion of such a confounder in the analysis would increase the estimate of the causal treatment effect from 0.142 to above 8, and also reduce its p-value from 0.877 to $< 0.05$.

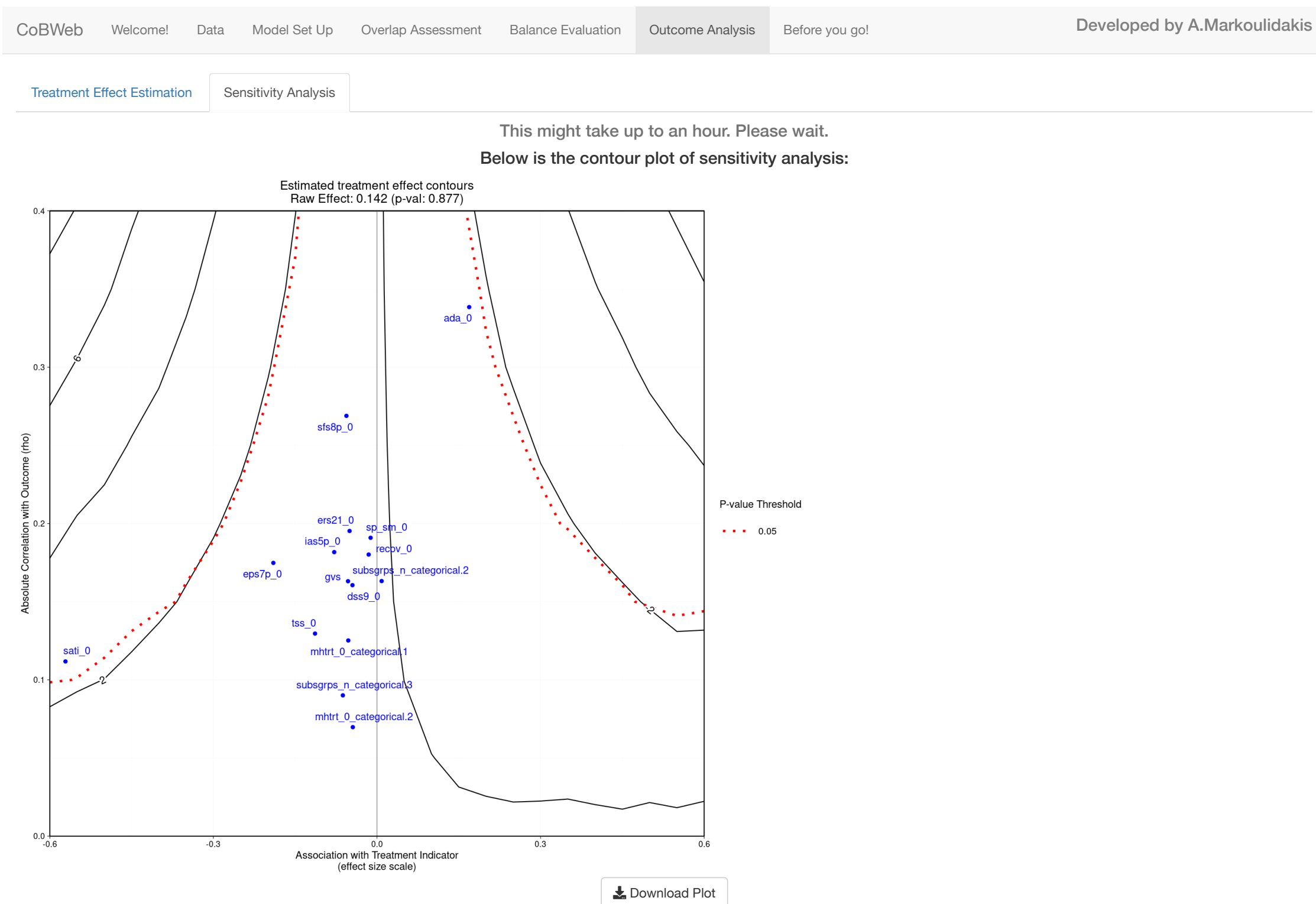

## *Before you go!*

The last module of the app ("Before you go!") includes options for the generation of a summary report of the analysis performed (in a .pdf document), and downloading the refined data (with the removal of observations with potential outliers) on .txt and/or .csv document(s), for further analysis.

# Discussion

Robust and unbiased estimation of causal treatment effect using data from observational studies is feasible if all steps of the analysis as outlined in this paper are carefully followed. To assist this process, we have created CoBWeb, a software tool intended for non-specialists (e.g. clinicians) to perform the key steps (Markoulidakis, Taiyari, et al., 2021) for the estimation of causal treatment effects using PS and balancing weights. One of the main advantages of the software is that a summary report in .pdf form is available to download, as well as the data (filtered for outliers) alongside the balancing weights, for further analysis.